\documentstyle[epsfig]{aipproc}

\begin{document}
\title{Hubble Space Telescope Imaging of the Field of
GRB970508}

\author{E. Pian$^1$, A. Fruchter$^2$,  L. E. Bergeron$^2$,
S. E. Thorsett$^3$,\\ F. Frontera$^{1,4}$, M. Tavani$^{5,6}$, E. Costa$^7$, 
M. Feroci$^7$,\\ J. Halpern$^6$, R. A. Lucas$^2$, L. Nicastro$^1$, 
E. Palazzi$^1$,\\ L. Piro$^7$, W. Sparks$^2$, A. J. 
Castro-Tirado$^8$, T. Gull$^9$,\\ K. Hurley$^{10}$, H.
Pedersen$^{11}$}
\address{$^1$ ITESRE-CNR, Via Gobetti 101, I-40129
Bologna, Italy\\ 
$^2$ STScI, 3700 San Martin Drive, Baltimore, MD 21218\\
$^3$ Joseph Henry Labs. and Dept. of Physics, Princeton 
University, Princeton, NJ 08544\\
$^4$ Dip. Fisica, Universit\`a di Ferrara, Via Paradiso 12, 
I-44100 Ferrara, Italy\\
$^5$ Columbia Astrophysics Laboratory, Columbia University, 
New York, NY 10027\\
$^6$ IFCTR-CNR, Via Bassini 15, I-20133 Milano, Italy\\
$^7$ IAS-CNR, Via E. Fermi 21, I-00044 Frascati, Italy\\
$^8$ Laboratorio de Astrof\'{\i}sica Espacial y F\'{\i}sica 
Fundamental, INTA, P.O. Box 50727, 28080 Madrid, Spain\\
$^9$ NASA/ Goddard Space Flight Center, Greenbelt, MD 20071\\
$^{10}$ University of California Space Sciences Laboratory, 
Berkeley, CA 94720\\
$^{11}$ Copenhagen University Observatory, Juliane Maries Vej 
30, DK 2100 Copenhagen, Denmark}

\maketitle

\begin{abstract}
We report on Hubble Space Telescope (HST) observations of the optical
transient (OT) discovered in the error box of GRB970508.  The object 
was imaged on 1997 June 2 with the Space Telescope 
Imaging Spectrograph (STIS) and Near-Infrared Camera and Multi-Object 
Spectrometer (NICMOS).  The observations reveal a point-like source with 
R = $23.1 \pm 0.2$ and H = $20.6 \pm 0.3$, in agreement with the power-law 
temporal decay seen in previous ground-based monitoring.  Unlike the case of 
GRB970228, no nebulosity is detected surrounding the OT of GRB970508,
although 
Mg I absorption and [O II] emission seen in Keck spectra at a redshift of
$z=0.835$ suggest the presence of a dense, star-forming medium.
The HST observations set very conservative upper limits of 
R $\sim 24.5$ and H $\sim 22.2$ 
on the brightness of any underlying extended source.  If this subtends a 
substantial fraction of an arcsecond, then the R band 
limit is $\sim$25.5.  
Subsequent photometry suggests a flattening of the light curve at
 later epochs.  Assuming the OT decline follows a pure power-law and
ascribing the flattening to the presence of an underlying
component of constant flux, we find that this must have
R = 25.4, consistent with the upper limits determined by HST.
At $z = 0.8$, this would correspond to an absolute magnitude
in the U band of $\sim -18$, similar to that of the Large Magellanic
Cloud (LMC).  We propose a scenario in which the host galaxy of the
GRB is of Magellanic type, possibly being a ``satellite'' of one of the
bright galaxies located at few arcseconds from the OT.

\end{abstract}

\section*{Introduction}

Optical and infrared imaging of GRB fields has been pursued
for several years with the aim of studying
the environments of the gamma-ray events, and
of localizing and characterizing their host sources.
However, till recently, due to the
insufficient angular precision in the knowledge of GRB
positions, the results have been inconclusive,
as reviewed by Band and Hartman\cite{bh97}.

The high positional accuracy attained by the BeppoSAX Wide Field
Cameras has allowed the pointing of optical telescopes at the
error circles of GRB970228 and 970508, and the discovery
of optical transients (OT) associated with them\cite{jvp97,bond97}.
For the former GRB, HST has resolved
an extended source surrounding
the point-like OT\cite{sahu97}, whose shape, colors and
constant brightness led to the identification with a galaxy of irregular
morphology.

\begin{figure}[b!] 
\centerline{\epsfig{file=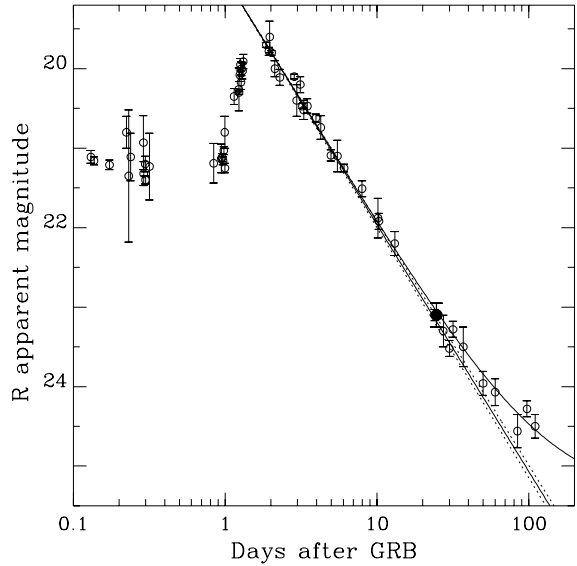,height=4.in,width=4.in}}
\vspace{10pt}
\caption{R band light curve of the OT associated to GRB970508. 
Photometry is from ground-based telescopes 
(open circles) and HST-STIS (filled circle). All magnitudes 
have been converted to 
Kron-Cousins R. Uncertainties have been rounded up to 0.1 magnitudes
when smaller values were reported in the literature, to take into
account possible systematic photometric offsets due to instrumental 
differences.  The fit with a power-law plus a constant with R = 25.4 is
reported (thick solid line) along with the power-law curve of index 
$-1.24$ (thin solid line) and its 1$\sigma$ uncertainty range (dashed
lines).}
\label{fig1}
\end{figure}

R band photometry of the OT of 
GRB970508\cite{h,a,s,djorg97,g,sc,so,m,gr,ga,c,metzger97b,j,gr2} 
shows that, following a first
increase, the flux started subsiding $\sim$2 days after the GRB
(see  Figure \ref{fig1}).
This decay can be modeled, till the epoch of the HST observation, 
with a  power-law temporal dependence of index 
$-1.17 \pm 0.04$.

Spectroscopy at the Keck II telescope\cite{metzger97a} reveals 
absorption systems at $z = 0.767$ and 0.835 superposed on the 
continuum as well as [O II] line emission at $z=0.835$\cite{metzger97b}.
These features, besides
proving unambiguously the extragalactic location of the OT,
suggest a line of sight through a dense interstellar medium; 
however, the only potential host galaxy detected from ground-based 
imaging is a faint blue object lying $5^{\prime\prime}.2$ away 
from the OT\cite{djorg97}. 

The HST observations presented here were designed to search for a 
host galaxy and to obtain late-time photometry of the OT of GRB970508. 
A detailed description of the data and results has been given
in Pian et al.\cite{pian98}, to which we refer for a complete
presentation.

\section*{Observations, data analysis and results}

Four 1250-second exposures were obtained of the GRB970508 field 
using the STIS CCD in Clear Filter mode during 1997 June 2.52-2.66 
(UT).  They have been dithered to allow removal of 
hot pixels and to obtain the highest possible resolution.  
The images 
were bias subtracted, flat-fielded, corrected for dark current and 
calibrated by the newly created STIS pipeline.  The final 
``drizzled''\cite{fh97} image
is available in the
Web (URL  http://www.stsci.edu/$\sim$fruchter/GRB/data\_970508).
Four exposures of 514 seconds each were also made with the NICMOS 
Camera 2 on 1997 June 2.67-2.74 (UT) and 
dithered using the NICMOS spiral dither pattern. The F160W filter 
(close to the standard near-infrared H band) was used.
The OT point-like source is easily visible in all of the data sets.

The photometric calibration of the images was done using the synthetic
photometry package SYNPHOT in IRAF/STSDAS. For STIS, given the broad-band 
response of the instrument, a power-law spectral shape with index 
$\alpha_\nu \simeq 1$ was assumed, based on Keck spectrophotometry and 
on the photometric colors. This yields for the OT V$=23.45 \pm 0.15$ 
($1\sigma$) and R$ = 23.10 \pm 0.15$.  
For NICMOS, our calibration gives an OT magnitude of H = $20.6 
\pm 0.3$. 
The faint galaxies located at North-East (G1, adopting the convention of
Djorgovski et al.\cite{djorg97}) and North-West of the OT (hereafter G2)
are found to have apparent magnitudes R = $24.8 \pm 0.2$, H = $22.8 \pm 0.1$
(G1), and R = $25.5 \pm 0.2$, H = $21.9 \pm 0.1$ (G2), respectively.

The R band STIS magnitude lies within the $1\sigma$ uncertainty of the 
extrapolation to June 2.5 of the power-law decay fit to earlier data.  An 
R band measurement (R = 23.4) taken at Keck after our HST observation 
(5 June) confirms this trend.
However, subsequent photometry suggests a slight flattening of the light
curve (see  Figure \ref{fig1}).  
Under the assumption that a power-law correctly reflects the behavior of
the OT, the flattening in the temporal descent might be the signature 
of an underlying component of constant brightness, such as a galaxy.
While a fit to the data with a simple power-law yields an index of
$ -1.14 \pm 0.02$ with a $\chi^2 = 1.6$, 
a fit with a power-law plus a constant gives a power-law 
best-fit index of $-1.24 \pm 0.02$, with $\chi^2 = 1.2$.
The fitted magnitude of the constant component is R = 25.4, with a 
90\% lower limit of  R = 24.7.

As visible from Figure 1, at the epoch of the HST observation it would
have not been possible to appreciate a deviation of the measured
flux from a simple power-law behavior. 
Consistently, direct inspection and analysis of the HST-STIS image
does not reveal any significant residual flux from an extended source 
underlying the OT either by subtracting  the STIS 
scaled stellar point spread function (PSF) from the image, nor by 
subtracting a ``compact galaxy'' PSF, i.e. the convolution of
a normal PSF with a Gaussian 
of intrinsic FWHM = $0^{\prime\prime}.15$.  This allows us to conclude
that any
underlying galaxy must be no brighter than R$=24.5$; if it is 
an extended object with a scale size greater than a few tenths of an 
arcsecond, it must be even fainter (R $\sim$25.5).  These limits are
consistent with the best fit value derived above.

Similarly, after subtraction of a scaled artificial PSF, the NICMOS 
image is also consistent with sky noise statistics and we estimate 
any underlying, extended component must have H$>22.2$ within 
$0^{\prime\prime}.4$ of the point-like source.  

\section*{Discussion}

The detection of Mg I absorption and [O II]
emission in the Keck spectra of the OT of GRB970508
implies that the absorbing medium  is not highly 
excited and that active star formation is occurring, respectively.  
While there are several galaxies with V$> 24.5$ within a few arseconds 
of the OT (e.g., G1 and G2), this corresponds to a projected distance of 
tens of kiloparsecs 
at $z=0.8$.  It seems unlikely that either the high density or low 
excitation necessary for the formation of the
Mg I line could be maintained this far out 
in a galactic halo\cite{ss92}.
Therefore, we believe that the absorbing medium 
responsible for these lines is presently hidden by the light from the 
OT and is almost certainly the underlying host galaxy.

The apparent magnitude of the constant component derived from the fit 
to the photometric points, dereddened with $A_R = 0.07$\cite{djorg97},
would correspond,  at $z = 0.8$, to an absolute
magnitude in the U band  of $-17.8$, with a 90\% lower limit of $-18.5$,
assuming $H_0 = 75$ km s$^{-1}$ Mpc$^{-1}$, and no $K$-correction.
This is consistent with the absolute de-extincted U
magnitude of the Large Magellanic Cloud (LMC), $-18$\cite{rc3}.
If a $K$-correction
of half a magnitude is assumed (appropriate for the optical-UV 
spectrum of the LMC\cite{rc3}), the absolute magnitude of the putative
host galaxy would be 
M$_U = -17.1$, and the absolute magnitude corresponding to the
90\% lower limit would be $-17.8$, still consistent 
with the luminosity of the LMC, given the uncertainties
in the cosmological parameters.
Therefore, it is possible that the GRB occurred in a small galaxy 
of brightness similar to that of the LMC and of comparable size
and shape.
Strong star formation takes place in Magellanic-like
galaxies\cite{hunter97}, so that
the proposed scenario would be still consistent with the speculation
of Pian et al.\cite{pian98} about the link of GRBs to star formation.

The OT is located at $\sim$5 arcseconds from the bright galaxies 
located on the North (G1 and G2).  If all three objects are at
$z = 0.8$, the GRB host galaxy would be at a few tens of
kiloparsecs away from those bright galaxies, and probably be
dynamically related to one of them.  Indeed,
G1 is extremely blue, as reported by Djorgovski et al.\cite{djorg97}, 
and the colors are consistent with a rapidly star-forming galaxy at any 
reasonable redshift.  G2 is somewhat redder, but has the colors of a 
nearby late-type spiral galaxy whose spectrum has been shifted to 
$z \sim 0.7-0.8$.  

In order to confirm or disprove this scenario, further optical and
near-infrared 
imaging of the OT of GRB970508 is necessary, which would
extend the sampling of the light curve and possibly
directly detect the hiding host galaxy.


\end{document}